\documentstyle[preprint,aps]{revtex}

\def\footstrut{\baselineskip 16pt}

\skip\footins = 14pt 
 \def\thebibliography#1{\vskip 1.5pc{\centerline {{\bf REFERENCES}}}\vskip 4pt
 \list
  {[\arabic{enumi}]}{\settowidth\labelwidth{[#1]}\leftmargin\labelwidth
  \advance\leftmargin\labelsep
  \usecounter{enumi}}
 \def\newblock{\hskip .11em plus .33em minus .07em}
  \sloppy\clubpenalty4000\widowpenalty4000
  \sfcode`\.=1000\relax}

\def\beq{\begin{equation}}
\def\eeq{\end{equation}}

\def\fnl{{\textstyle{\frac{N}{L}}}}
\def\fnll{{\textstyle{\frac{N}{L^2}}}}

\begin{document}
\tighten

\nonfrenchspacing
\flushbottom
\title{A Limit on the Speed of Quantum  Computation\\  for  Insertion into an
Ordered List\thanks{\baselineskip=16pt This work was supported in part by The Department of
Energy under cooperative agreement DE-FC02-94ER40818 and by the National Science
Foundation under grant NSF 95--03322 CCR.}}

\def\footstrut{\baselineskip 12pt}

\author{Edward Farhi\thanks{farhi@mit.edu} and Jeffrey
Goldstone\thanks{goldstone@mitlns.mit.edu}}
\address{Center for Theoretical Physics \\ 
Massachusetts Institute of Technology \\
Cambridge, MA  02139}

\author{Sam Gutmann\thanks{sgutm@nuhub.neu.edu}} 

\address{Department of Mathematics \\ 
Northeastern University \\ 
Boston, MA 02115}

\author{Michael Sipser\thanks{sipser@math.mit.edu}}

\address{Department of Mathematics \\ 
Massachusetts Institute of Technology \\
Cambridge, MA  02139 \\[2em]
{\rm MIT-CTP-2811~~quant-ph/9812057 \hfil \qquad \qquad December1998}}
\maketitle

\setcounter{page}{0}
\thispagestyle{empty}

\vspace{.5in}

\begin{abstract}

We consider the problem of inserting a new item into an ordered list of N-1 items.  The length
of an algorithm is measured by the number of comparisons it makes between the new item
and items already on the list.  Classically, determining the insertion point requires log N
comparisons.  We show that, for N large, no quantum algorithm can reduce the number of
comparisons below log N/(2 loglog N).
\end{abstract}


\section{Introduction}

Quantum mechanical algorithms can outperform classical algorithms in certain cases.  Classically
searching an unordered list of $N$ items takes of order $N$ queries.  Quantum mechanically this
can be accomplished with of order $\sqrt{N}$ queries \cite{ref:1}.  There is also a lower bound
of $\sqrt{N}$ for this quantum problem \cite{ref:2} so the square root speedup is optimal.
However square root speedup is not universal.  For example, determining the parity of a list of
$N$ items, each equal to plus one or minus one, requires $N$ queries classically and at least
$N/2$ queries quantum mechanically \cite{ref:3},\cite{ref:4}.

In this paper, we consider the problem of inserting a new item into an ordered list of $N-1$
 items.  A single (classical) query consists of comparing the new item with any chosen
item on the list to see if the new item comes before or after the chosen item.  Classically, the best
algorithm for determining the point of insertion is binary search, which uses $\log_2 N$ queries. 
We show that quantum mechanically, for large
$N$,  an algorithm which succeeds after $k$ quantum queries must have
\beq
k>\frac{\log_2 N}{2 \log_2 \log_2 N} \ \ .
\label{eq:new1}
\eeq
 The same bound is
obtained if we only ask the algorithm to determine the point of insertion correctly with
probability
$\epsilon>0$ (where $\epsilon$ does not depend on $N$). A lower bound of order
$\sqrt{\log N}/ \log \log N$ recently appeared in \cite{ref:5}. 

Our result shows that possible
quantum improvement in this problem is at most modest.  In a sequel to this paper we will
demonstrate a quantum algorithm that succeeds in $c \log_2 N$ quantum queries with $c<1$. 

\section{Preliminaries}

The classical problem of inserting one item into an ordered list of $N-1$ items is equivalent to the
following oracular problem.  Consider the $N$ functions $F_j$ defined on the set $\{ 1,2,\dots
N\}$ by
\beq
F_j(x) = \left\{
\begin{array}{cc}
1 & x<j \\[1ex]
0 & x \ge j
\end{array}
\right.
\label{eq:1}
\eeq
for $j=1,2,\dots N$.  A query consists of giving the oracle a value of $x$ with the oracle returning
$F_j(x)$ for some fixed but unknown $j$.  The problem is to determine $j$.  Binary search
determines $j$ with $\log_2 N$ queries, which is the optimal result classically.

(The reader may have noticed that $F_j (N)=1$ for all $j$ so querying the oracle with $x=N$ is of
no help.  However for later convenience we want the size of the domain of $F_j$ to equal the
number of functions.)

Quantum mechanically we work in a Hilbert space with an orthonormal basis
\beq
\{ |x,q,w\rangle \} ~{\rm with}~
\left\{
\begin{array}{rcl}
x &=& 1,2,\dots N \\[1ex]
q &=& 0,1 \\[1ex]
w &=& 1,2,\dots W 
\end{array}
\right.
\label{eq:2}
\eeq
where $q$ and $w$ label a basis for the work-space of dimension $2W$.  Given an oracle
associated with any function $F(x)$ which can take the values 0 and 1, a quantum query is an
application of the unitary operator, $\hat{F}$, defined by
\beq
\hat{F} |x,q,w\rangle = |x,F(x) \oplus q, w \rangle \ \ .
\label{eq:3}
\eeq
A quantum algorithm which makes $k$ queries starts with a state $|s\rangle$ (which is a
superposition of the states in (\ref{eq:2})) and alternately applies $\hat{F}$ and
$F$-independent unitary operators, $V_i$, producing
\beq
 |\psi_F\rangle = V_k \hat{F} V_{k-1} \cdots \hat{F} V_1 \hat{F} |s\rangle \ \ .
\label{eq:4}
\eeq
The $V$'s may act in the full Hilbert space.  Designing an algorithm consists of choosing 
$|s\rangle$ and the $V$'s.  For the insert problem where $F(x)$ is guaranteed to be one of the
$N$ functions $F_j(x)$ of the form (\ref{eq:1}), a $k$-query algorithm succeeds if
$|\psi_{F_\ell}\rangle$ is orthogonal to $|\psi_{F_m}\rangle$ for $\ell \ne m$.  A single
measurement will then distinguish the $N$ different $F_j$.

\section{Main Result}

We now show that, for $N$ large,  a $k$-query algorithm cannot distinguish the $N$
different
$F_j$ if $k$ is less than $\frac{\log_2 N}{2 \log_2 \log_2 N}$.
Consider a particular $k$-query algorithm of the form
(\ref{eq:4}), that is, a sequence of queries alternating with other unitary transformations, acting on a
fixed initial state.  A successful algorithm achieves an adequate separation of the final states 
associated with the different values of $j$.  However the separation obtained with a single query
is limited.  We show that after the first query there is a range of consecutive $j$'s for which the
algorithm has achieved little separation.  Similarly after the second query there is a subrange of
consecutive
$j$'s for which the algorithm has achieved only slightly more separation.  We can continue
subdividing the range of
$j$'s with each subsequent query as long as at least two values of $j$ remain in the range.  If, in
fact, after $k$ queries, two values of $j$ remain and the separation between the two
corresponding states is still small, then the algorithm has failed.

To carry this out we first we define projectors
\begin{mathletters}
\beq
P(a,b) = \sum_{a <x \le b} \sum_q \sum_w |x,q,w \rangle \langle x,q,w|
\label{eq:6a}
\eeq
with $0\le a<b \le N$.  Note that these operators act as the identity on the work-space.  The
orthogonal projectors are given by 
\beq
P^\bot (a,b) = \bigg(\sum_{x\le a} + \sum_{x>b} \bigg)\sum_q  \sum_w |x,q,w \rangle
\langle x,q,w|
\label{eq:6b}
\eeq%
\label{eq:6}%
\end{mathletters}%
and clearly $P (a,b) +P^\bot (a,b)$ is the identity operator.

The algorithm starts in the state $|s\rangle$.  We write $|s\rangle$ as a sum of $L$ orthogonal pieces
\beq
|s\rangle = \sum^{L-1}_{r=0} P \Big(r \fnl, (r+1)\fnl
\Big) |s\rangle
\label{eq:7}
\eeq
where $L$ is to be determined later but is small compared to $N$.  (For clarity, we
pretend that certain numbers such as $\frac{N}{L}$ are always integers.)  There is an $r$, call it
$r_1$, such that
\beq
\Big|\Big|  P \Big(r_1 \fnl, (r_1 +1)\fnl \Big) |s\rangle \Big|\Big|  \le
\frac{1}{\sqrt{L}} \ \  .
\label{eq:8}
\eeq
We now write
\begin{mathletters}
\begin{eqnarray}
|s\rangle &=& | \phi_1\rangle + | \phi_1^\bot \rangle
\label{eq:9a} \\
\noalign{\hbox{where}}
| \phi_1\rangle &=& P \Big(r_1 \fnl, (r_1 +1)\fnl \Big) |s\rangle
\label{eq:9b} \\
\noalign{\hbox{and}}
| \phi_1^\bot\rangle &=& P^\bot \Big(r_1 \fnl, (r_1 +1)\fnl \Big) |s\rangle
\label{eq:9c} \ \ .
\end{eqnarray}
\label{eq:9}%
\end{mathletters}%
We assume that the oracle holds the function $F_j(x)$.  After one quantum query and the
application of the first $F$-independent unitary operator we arrive at 
\beq
V_1 \hat{F}_j |s\rangle = V_1 \hat{F}_j | \phi_1\rangle + V_1 \hat{F}_j | \phi_1^\bot\rangle \ \ .
\label{eq:10}
\eeq
Observe that $V_1 \hat{F}_j | \phi_1^\bot\rangle$ does not vary with $j$ for $r_1 \frac{N}{L}<j
\le  (r_1 +1) \frac{N}{L}$ and for these $j$ we write
\beq
V_1 \hat{F}_j | \phi_1^\bot\rangle = |\gamma_1 \rangle \ \ .
\label{eq:11}
\eeq
We now write $P \Big(r_1 \frac{N}{L}, (r_1+1) \frac{N}{L} \Big) |\gamma_1 \rangle$, whose norm
is at most 1, as the sum of $L$ orthogonal pieces
\beq
P \Big(r_1 \fnl, (r_1+1) \fnl \Big) |\gamma_1 \rangle =
\sum^{L-1}_{r=0}  P\Big(r_1 \fnl + r \fnll, r_1 \fnl + (r+1) \fnll \Big) |\gamma_1 \rangle 
\label{eq:12}
\eeq
and choose a value of $r$, call it $r_2$, such that
\beq
\Big|\Big| P \Big(r_1 \fnl+ r_2 \fnll, r_1 \fnl + (r_2+1)
\fnll \Big) |\gamma_1 \rangle \Big|\Big| \le \frac{1}{\sqrt{L}} \ \ .
\label{eq:13}
\eeq
We write $|\gamma_1 \rangle  = |\phi_2 \rangle +  |\phi_2^\bot \rangle$ with 
\begin{mathletters}
\begin{eqnarray}
 |\phi_2 \rangle &=& P \Big(r_1 \fnl + r_2 \fnll, r_1 \fnl + (r_2+1)
\fnll \Big) |\gamma_1 \rangle
\label{eq:14a} \\
\noalign{\hbox{and}}
 |\phi_2^\bot \rangle &=& P^\bot \Big(r_1 \fnl+ r_2 \fnll, r_1 \fnl +
(r_2+1)
\fnll \Big) |\gamma_1 \rangle \ \ .
\label{eq:14b}
\end{eqnarray}
\label{eq:14}%
\end{mathletters}%
We now have 
\beq
V_1 \hat{F}_j | s\rangle =V_1 \hat{F}_j |\phi_1 \rangle +  |\phi_2 \rangle +  |\phi_2^\bot \rangle
\ \ .
\label{eq:15}
\eeq
Application of $\hat{F}_j$ and $V_2$ yields
\beq
V_2 \hat{F}_j V_1 \hat{F}_j| s\rangle = V_2 \hat{F}_j V_1 \hat{F}_j |\phi_1 \rangle + V_2
\hat{F}_j |\phi_2 \rangle +  V_2 \hat{F}_j |\phi_2^\bot \rangle \ \ .
\label{eq:16}
\eeq
The state $V_2 \hat{F}_j |\phi_2^\bot \rangle$ does not vary with $j$ for the $\frac{N}{L^2}$
values of $j$ in the range $r_1 \frac{N}{L}+ r_2 \frac{N}{L^2} <j \le r_1 \frac{N}{L}+(r_2+1)
\frac{N}{L^2}$.  Recall that $\Big|\Big| \, |\phi_1\rangle \Big|\Big|\le \frac{1}{\sqrt{L}}$ and
$\Big|\Big| \, |\phi_2\rangle \Big|\Big|
\le \frac{1}{\sqrt{L}}$ so by unitarity
\begin{eqnarray}
\Big|\Big| V_2 \hat{F}_j V_1 \hat{F}_j |\phi_1\rangle \Big|\Big| \le \frac{1}{\sqrt{L}}
\nonumber \\
\noalign{\hbox{and}}
\Big|\Big| V_2 \hat{F}_j  |\phi_2\rangle \Big|\Big| \le \frac{1}{\sqrt{L}} \ \ .
\label{eq:17}
\end{eqnarray}
Repeating this procedure for a total of $k$ queries and correspondingly $k$ subdivisions, we get
\beq
V_k \hat{F}_j \cdots V_1 \hat{F}_j| s\rangle = |\delta_1 \rangle + |\delta_2 \rangle + \dots
|\delta_k \rangle + V_k \hat{F}_j |\phi_k^\bot \rangle
\label{eq:18}
\eeq
where $\Big|\Big| \, |\delta_i \rangle \Big|\Big| \le \frac{1}{\sqrt{L}}$ for $i=1,\dots k$ and $V_k
\hat{F}_j |\phi_k^\bot\rangle$ does not vary with $j$ in a range of size $\frac{N}{L^k}$. 
Eq.~(\ref{eq:18}) can also be written as 
\beq
V_k \hat{F}_j \cdots V_1 \hat{F}_j| s\rangle = |\delta \rangle + V_k \hat{F}_j |\phi_k^\bot \rangle
\label{eq:19}
\eeq
where 
$$
|\delta \rangle = \sum_i |\delta_i \rangle
$$
and
\beq
\Big|\Big| \, |\delta \rangle \Big|\Big| \le \frac{k}{\sqrt{L}} \ \ .
\label{eq:20}
\eeq
Note that $|\delta \rangle$ in general depends on $j$ although this is not explicitly indicated.

For the $k$-query algorithm to be successful, the states on the left-hand side of (\ref{eq:18})
must be orthogonal for $j \ne j'$.  Suppose $L$ is chosen such that $\frac{N}{L^k} \ge 2$. 
This implies that there exist two values of $j$, say $j'$ and $j''$ for which the states
\beq
|\delta' \rangle + V_k \hat{F}_{j'}  |\phi_k^\bot \rangle \quad {\rm and} \quad
|\delta^{\prime\prime} \rangle + V_k \hat{F}_{j''}   |\phi_k^\bot \rangle
\label{eq:21}
\eeq
are orthogonal but
\beq
V_k \hat{F}_{j'}  |\phi_k^\bot \rangle =
V_k \hat{F}_{j''}   |\phi_k^\bot \rangle \ \ .
\label{eq:22}
\eeq
Taking the difference of the two states in (\ref{eq:21}) gives
\beq
\Big|\Big| \, |\delta' \rangle - |\delta'' \rangle \Big|\Big| = \sqrt{2}
\label{eq:23}
\eeq
since the states in (\ref{eq:21}) have unit norm.
However by (\ref{eq:20}) we see that (\ref{eq:23}) is impossible if 
\beq
\frac{k}{\sqrt{L}} < \frac{1}{\sqrt{2}}
\label{eq:24}
\eeq
Thus our $k$-query algorithm cannot succeed if there exists an $L$ such that  (\ref{eq:24}) is
true and 
\beq
\frac{N}{L^k} \ge 2 \ \ .
\label{eq:25}
\eeq
This implies that, for large $N$, no $k$-query algorithm can succeed unless
\beq
2^{k+1} k^{2k} \ge N \ \ .
\label{eq:26}
\eeq

For an algorithm that determines the correct point of insertion with probability
$\epsilon>0$ ($\epsilon$~independent of $N$),  the states in (\ref{eq:21}) must have
the absolute value of their inner product at most  $\epsilon'$ where $\epsilon'$ depends on
$\epsilon$.  This is impossible if $\frac{k}{\sqrt{L}}< \rho$, for some $\rho$ depending on
$\epsilon'$.  In this case, (\ref{eq:26}) becomes
\beq
2 \Big( \frac{1}{\rho} \Big)^{2k} k^{2k} \ge N \ \ .
\label{eq:27}
\eeq
For large $N$, either (\ref{eq:26}) or (\ref{eq:27}) requires that condition (\ref{eq:new1}) holds.

\section{Remark}

What is quantum mechanical about this proof?  If the unitary operators were replaced by
stochastic matrices, and $\Big|\Big| \cdot \Big|\Big|$ reinterpreted as the $L_1$ norm instead of
the $L_2$ norm, then the $\frac{1}{\sqrt{L}}$, in (\ref{eq:8}) and succeeding formulas, becomes
$\frac{1}{L}$, and the bound ultimately becomes $\frac{\log_2 N}{\log_2\log_2 N}$.  Only the
factor of 2 changes.  Of course this is not the best classical lower bound, which is $\log_2 N$.



\begin{thebibliography}{Ozhigov}

\bibitem{ref:1} L.K. Grover, ``A Fast Quantum Mechanical Algorithm for Database
Search,'' quant-ph/9605043.

\bibitem{ref:2} C.H. Bennett, E. Bernstein, G. Brassard and U.V. Vazirani, ``Strengths and
Weaknesses of Quantum Computing,'' quant-ph/9701001.

\bibitem{ref:3} E. Farhi, J. Goldstone, S. Gutmann, M. Sipser, ``A Limit on the Speed of Quantum
Computation in Determining Parity,'' quant-ph/9802045.

\bibitem{ref:4} R. Beals, H. Buhrman, R. Cleve, M. Mosca, R. deWolf, ``Quantum Lower Bounds by
Polynomials,'' quant-ph/9802049.

\bibitem{ref:5} H. Buhrman, R. deWolf, ``Lower Bounds for Quantum Search and
Derandomization,'' quant-ph/9811046.
 

\end{thebibliography}
\end{document}